\documentclass[traditabstract]{aa}
\usepackage{graphicx}
\usepackage{txfonts}
\usepackage{rotating}
\usepackage{lscape}
\usepackage{longtable}
\usepackage{natbib}

\begin{document}

\title{IGR J17488--2338: a newly discovered  giant radio galaxy}
\titlerunning{ IGR J17488--2338: a newly discovered  giant radio galaxy}
\authorrunning{M.~Molina}
\author{M. Molina\inst{1} \and L. Bassani\inst{1} \and A. Malizia\inst{1} 
\and A.~J. Bird\inst{2} \and A. Bazzano\inst{3} \and P. Ubertini\inst{3} \and T. Venturi\inst{4} 
 }
 
\offprints{molina@iasfbo.inaf.it}
\institute{IASF/INAF, via Gobetti 101, I-40129 Bologna, Italy \and
School of Physics and Astronomy, University of Southampton,
        SO17 1BJ, Southampton, U.K. \and IAPS/INAF Via Fosso del Cavaliere 100, I-00133 Roma \and 
        IRA/INAF, via Gobetti 101, I-40129 Bologna }

\date{Received  / accepted}

\abstract
{We present the discovery of a large scale radio structure associated with IGR J17488--2338, 
a source recently discovered by \emph{INTEGRAL}
and optically identified as a broad line AGN at redshift 0.24. At low frequencies, the source  
properties are those of an intermediate-power FR II radio galaxy with a linear size of 1.4\,Mpc. 
This new active galaxy is therefore a member of a class of objects called Giant Radio Galaxies (GRGs),
a rare type of radio galaxies with physical sizes larger than 0.7\,Mpc; they represent the largest 
and most energetic single entities in the Universe and are useful laboratories for many astrophysical
studies. Their large scale structures could be due either to special external conditions 
or to uncommon internal properties of the source central engine 
The AGN at the centre of IGR J17488--2338  
has a black hole of 1.3$\times$10$^9$ solar masses, a bolometric luminosity of 
7$\times$10$^{46}$erg\,s$^{-1}$ and 
an Eddington ratio of 0.3, suggesting that it is powerful enough to produce the large structure
observed in radio. The source is remarkable also for other properties, among which its X-ray
absorption, at odds with its type 1 classification, and the presence of a strong iron line which 
is a feature not often observed in radio galaxies.  
 
\keywords{gamma-rays: galaxies - radio continuum galaxies - galaxies: active }
}

\maketitle

\section{Introduction}

Powerful extragalactic radio sources are galaxies (and/or quasars) hosting active galactic nuclei
(AGNs), which produce jets and extended radio emitting regions (lobes) of plasma. Some of them are
characterised by giant structures and are known as giant radio galaxies (GRG), formally those with 
linear sizes larger than 0.7\,Mpc (e.g. \citealt{lara:2001,Ishwara-Chandra:1999}, scaled 
for the cosmology adopted here of H$_0$= 71 km\,s$^{-1}$\,Mpc$^{-1}$, $\Omega_{\rm m}$=0.27,
$\Omega_{\Lambda}$=0.73). These objects represent the largest and most energetic single entities 
in the Universe and it is possible that they play a special role in the formation of large-scale 
structures. They generally belong to the FR II \citep{fanaroff:1974} radio morphology 
(edge brightened), have relatively low radio power 
(LogP$_{\rm 1.4GHz (W/Hz)}$$\gtrsim$24.5; \citealt{owen:1989}) and reside in elliptical 
galaxies and quasars.
GRGs are very useful for studying many astrophysical issues, such as understanding the evolution of
radio sources, probing the intergalactic medium at different redshifts, investigating the nature of 
their central AGN. There are various scenarios which try to explain this 
phenomenon. For example, GRG could be very old sources which had had enough time to evolve to 
such large sizes. Alternatively, they could grow in an intergalactic medium whose density is 
smaller than that surrounding smaller radio sources, or, instead, their AGNs are extremely powerful 
and/or long-lived and thus able to produce such large scale structures. 
Because of their large sizes and relatively low 
radio power, the surface brightness of GRGs is low. This is why they are so 
difficult to find even in radio surveys and why the finding of a new member of the class is 
interesting and useful. Here we report on the discovery and subsequent analysis of the 
\emph{INTEGRAL} source IGR J17488--2338, which we have identified as an FR II radio galaxy with a 
linear size of 1.4\,Mpc, thus well above the threshold for it to be a GRG.  
In optical terms, the galaxy is a Seyfert 1.5 at redshift 0.24 and the source is 
remarkable also for other properties, among which its absorption characteristics, at odds with its 
type 1 classification, and the presence of an iron line in the X-rays. In particular, the core of this 
giant radio galaxy is extremely powerful in X/gamma-rays, is highly massive and able to accrete very 
efficiently: it is possible that these extreme properties provide the necessary conditions (high jet 
power or long activity time) to produce the giant radio structure observed in this newly discovered 
radio galaxy.

\section{Source discovery and identification}

IGR J17488--2338 was first reported by \citet{Bird:2010} as an unidentified and faint X-ray source 
detected by \emph{IBIS} during observations of the Galactic plane. 
Here we use data collected in the 4$^{\rm th}$ \emph{IBIS} survey \citep{Bird:2010}, which 
consists of all exposures from the beginning of the mission (November 2002) up to April 2008. The 
total exposure on this region is $\sim$6.2\,Ms. \emph{IBIS/ISGRI} images for each available pointing 
were generated in various energy bands using the ISDC offline scientific analysis software version 7.0 
(OSA 7.0; \citealt{Goldwurm:2003}) and the light curves generated with this method
where analysed first at revolution level and then applying the bursticity analysis. The source was 
clearly detected at the revolution level, but barely over the entire observation period with an upper 
limit on the average 20--40\,keV flux of 0.2\,mCrab and a 4$\sigma$ detection of 
0.4$\pm$0.1\,mCrab in the 40--100\,keV band.
The source has a bursticity of 5.3 suggesting significant variability. The peak flux, defined as the 
mean flux during the single period of time over which the significance is maximized (roughly 2.6 days 
around October 13--15, 2003) is 3.4$\pm$0.8\,mCrab in the 20--40\,keV band; the source dynamic range 
is therefore at least a factor of 17.
The source spectrum corresponding to this period of high flux level is well fitted by a power law with 
$\Gamma$=1.30$\pm$0.35 and a 20--100\,keV flux of 1.40$\times$10$^{-11}$erg\,cm$^2$\,s$^{-1}$. 
  
IGR J17488--2338 is located at R.A.(J2000) = 17h 48m 47.3s and
DEC(J2000) = -23d 38m 06.0s, with a positional uncertainty of $\sim$4.7$^{\prime}$.
The source has not yet been reported in the latest \emph{Swift/BAT} 
surveys (\citealt{Baumgartner:2013}, \citealt{Cusumano:2010}); note, however,
that these surveys mainly look for persistent sources.

\citet{stephen10}, by means of a cross-correlation analysis with the \emph{XMM} Slew Survey Catalogue 
\citep{saxton:2008}, reported the first X-ray counterpart of IGR J17488--2338 in a source 
(XMMSL1 J174838.8--233527) located at R.A.(J2000) = 17h 48m 38.9s and
DEC(J2000) = -23d 35m 26.8s (with an error radius of 5.1$^{\prime\prime}$) and displaying a 2--12\,keV 
flux of 2.5$\times$10$^{-12}$erg\,cm$^2$\,s$^{-1}$.
In subsequent X-ray follow-up observations with \emph{Swift/XRT}, \citet{landi:2011} confirmed this 
association and measured the source broad-band X-ray spectrum for the first time: over the 
0.3--100\,keV band IGR J17488--2338 displays a power-law continuum with a flat spectral index 
($\Gamma$=1.3), absorption in excess to the Galactic value 
(N$_{\rm H}$=1.3$\times$10$^{23}$\,cm$^{-2}$) and a 2--100\,keV flux of 
$\sim$2$\times$10$^{-11}$erg\,cm$^2$\,s$^{-1}$  
Despite being classified as a variable object in the fourth \emph{IBIS} 
catalogue \citep{Bird:2010}, the cross-calibration constant between \emph{INTEGRAL/IBIS} and 
\emph{Swift/XRT} turned out to be compatible with unity; similarly, the source 2--10\,keV flux was 
found to be fully consistent with the \emph{XMM} Slew one.

The reduction in the source positional uncertainty made possible by the identification of the X-ray 
counterpart allowed further radio and optical follow-up studies. Firstly, the X-ray source was found 
to coincide with the central region of a bright radio source, showing 
a double lobe morphology \citep{stephen10}; this association immediately suggested an extragalactic 
nature for the source, i.e. the discovery of a new radio galaxy with an active nucleus located behind 
the Galactic plane. IGR J17488--2338 was then optically classified by \citet{Masetti:2013} as a 
Seyfert 1.5 galaxy at z = 0.24. These authors also measured the mass of the source central black hole 
using the broad line velocities; for IGR J17488--2338 the H$\beta$ emission line provides a 
black hole of 1.3$\times$10$^9$ solar masses.

\begin{figure}
\begin{center}
\includegraphics[scale=0.45]{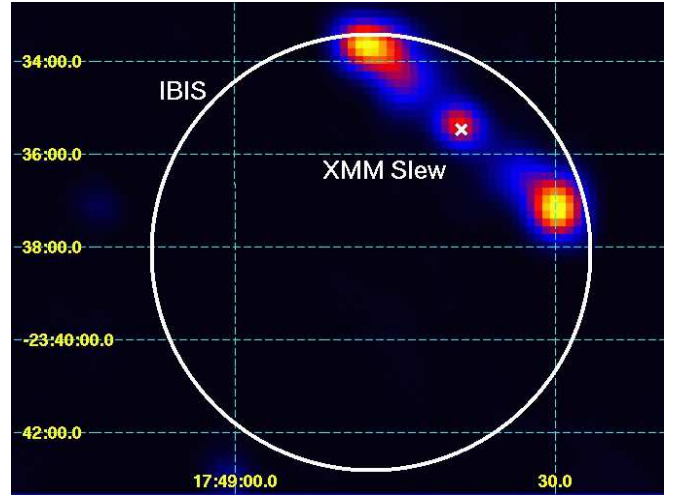}
\caption{1.4\,GHz NVSS image of IGR J17488--2338. The white circle corresponds to the
\emph{IBIS} error circle, while the white cross marks the XMM Slew position.}
\label{atel}
\end{center}
\end{figure}

Figure~\ref{atel} shows a cut-off image from the 1.4\,GHz NVSS survey \citep{condon98} 
of the region surrounding IGR J17488--2338; the figure shows the clear presence of a 
double-lobe radio source inside the \emph{INTEGRAL} positional uncertainty (white large circle) with 
the cross marking the position of the X-ray source detected by the \emph{XMM} Slew survey. 
The two radio lobes are fairly symmetrical, roughly ellipsoidal and placed on either side of the 
compact X-ray source; they are most likely powered by two symmetrically-positioned jets 
emanating from the AGN core which is marked by the X-ray emission.

\section{Further information on the source}

\begin{figure}
\begin{center}
\includegraphics[scale=0.75]{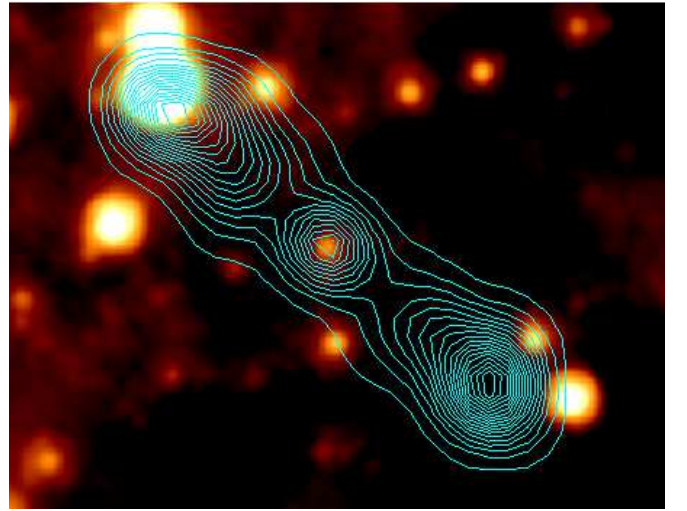}
\caption{22.1 micron WISE image of the region containing IGR J17488--2338; NVSS radio contours are 
superimposed.}
\label{wise}
\end{center}
\end{figure}

The X-ray counterpart of IGR J17488--2338 is listed in the WISE survey \citep{wright10} 
with 3.5, 4.6, 11.6 and 22.1 micron magnitudes of 
12.12, 11.02, 8.5 and 5.9 respectively; at the source redshift the estimated WISE luminosity
is of the order of 3$\times$10$^{45}$erg\,s$^{-1}$. The first two magnitudes are reported to be 
variable with a 
high probability. The 22.1 micron WISE image of the region around the source is shown
in Figure~\ref{wise} overlaid with the 1.4\,GHz contours.
The WISE object is clearly associated to the core of the radio galaxy, while no extended infrared 
emission appears to be produced in correspondence of the lobes (the bright spot associated to one of 
them is likely a background/foreground object); this is consistent with 
the fact that no significant thermal emission is expected to be associated with the lobes of radio 
galaxies.  

The radio core of IGR J17488--2338 is also reported as a 2MASS object with J, H, K fluxes ranging 
from 14 to 13 magnitudes, providing a near infrared luminosity around 
4$\times$10$^{45}$erg\,s$^{-1}$. In the optical the source has 
R and B magnitudes of 17.2 and 19 respectively, yielding a lumonosity of 
2$\times$10$^{45}$erg\,s$^{-1}$. The source is quite red (R--K = 4.3),
possibly due to heavy obscuration along the line of sight related to the source location close to the 
galactic plane; indeed \citet{Masetti:2013} measure 4.9 magnitudes of extinction due to our galaxy 
alone.

To improve the quality of the source X-ray spectrum, we also acquired an \emph{XMM-Newton} observation 
which was performed on September 9, 2013 during revolution 251. The EPIC-pn image of the region
shows two well detected objects: the one already seen by \emph{Swift/XRT} and by the \emph{XMM}
Slew survey, plus an extra source located at R.A.(J2000) = 17h 48m 49.49s and
DEC(J2000) = -23d 37m 39.8s. This second source is however very soft (undetected above few keV) 
and unlikely to be an alternative/valid counterpart of IGR J17488--2338: it is instead likely 
associated to coronal emission from the star HD 161837 of G5V spectral type. 
Therefore, also \emph{XMM-Newton} confirms the association with the radio source and 
further indicates that no X-ray emission is detected from the lobes but only from the central radio 
core. 

EPIC-pn \citep{Turner:2001} data were reprocessed using the \emph{XMM-Newton} Standard
Analysis Software (SAS) version 12.0.1 and employing the latest available calibration files. 
Only patterns corresponding to single and double events (PATTERN$\leq$4) were taken
into account; the standard selection filter FLAG=0 was applied.  
The EPIC-pn nominal exposure of 25\,ks has been filtered for periods of high background and the
resulting exposure amounts to $\sim$14\,ks. Source counts were extracted from 
a circular region of 23$^{\prime\prime}$ radius centred on the source, 
while background spectra were extracted from two circular regions of 20$^{\prime\prime}$ of radius 
close to the source. 
The ancillary response matrix (ARF) and the detector response matrix (RMF) were
generated using the \emph{XMM}-SAS tasks \emph{arfgen} and \emph{rmfgen}
and spectral channels were rebinned in order to achieve a minimum of 20 counts per each bin.
No pile-up was detected for this source. The spectral analysis was performed 
using \texttt{XSPEC} v.12.8.0 \citep{Arnaud:1996}; errors are 
quoted at 90\% confidence level for one parameter of interest 
($\Delta\chi^2$=2.71). As done previously, we combine the soft X-ray data  data obtained from the 
EPIC-pn observation to the \emph{IBIS/ISGRI} points obtained during the source outburst; to take into 
account any mis-match between these two instruments and/or any flux variability 
between observations a cross-calibration constant was introduced 
in the fit. 

\begin{figure}
\begin{center}
\includegraphics[scale=0.35, angle=-90]{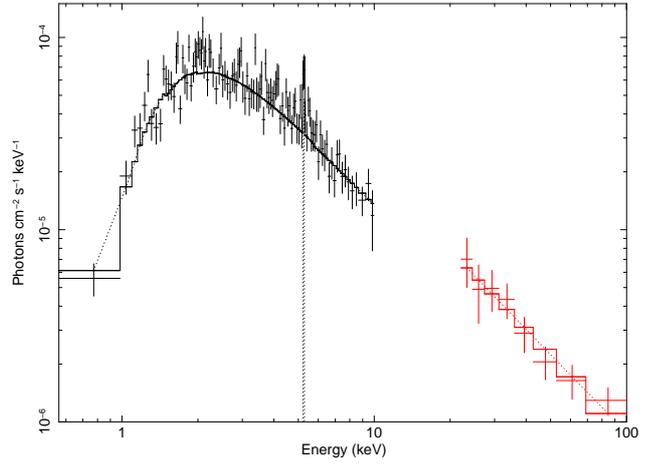}
\caption{0.5-110\,keV XMM-IBIS best fit plot; black data points are XMM measuremnts, while red data points are the IBIS one. The model employed is a partially absorbed power law plus a neutral iron line.}
\label{po}
\end{center}
\end{figure}

The 0.5--110\,keV broad-band spectrum was first fitted using a simple power law absorbed only by  
Galactic column density, which in the source direction is N$_{\rm H}$=0.41$\times$10$^{22}$cm$^{-2}$; 
the fit is quite poor, yielding a $\chi^2$=575.4 for 321 d.o.f. and a flat power law photon
index ($\Gamma$=0.61$\pm$0.04). In the model-to-data ratios, clear residuals are present at 
low energies (probably due to absorption in excess of the Galactic value) and around 
the iron K$\alpha$ line (which in the observer's rest frame is around 5.5\,keV).
The addition of intrinsic absorption improves the fit significantly 
($\chi^2$=313.9 for 320 d.o.f.) but maintains some residuals below 1\,keV; the resulting
column density is N$_{\rm H}$=0.91$^{+0.15}_{-0.14}$$\times$10$^{22}$cm$^{-2}$. 
To achieve a better fit, we substitute the fully covering cold absorber with a partially covering 
model. The fit further improves ($\chi^2$=307.9 for 319 d.o.f.) providing a column density  
N$_{\rm H}$=1.14$^{+0.26}_{-0.23}$$\times$10$^{22}$cm$^{-2}$
covering 93\% of the central source. We then added a Gaussian component, with width 
fixed to 10\,eV, to model the iron line. This model provides further improvement in the fit 
($\chi^2$=297.3 for 317 d.o.f.); the spectral slope is now $\Gamma$=1.37$\pm$0.11 and the 
iron line is found at E$_{\rm K\alpha}$=6.51$\pm$0.07\,keV (source rest frame) with
an equivalent width of 128$^{+61}_{-62}$\,eV. This is the best fit model
which is also displayed in Figure~\ref{po}. In fact, adopting a more complex model like  
an exponentially cut-off power law reflected from neutral material, the fit does not improve 
($\chi^2$=297.36 for 315 d.o.f) and only a loose constraint on the reflection fraction can
be obtained (R$<$1.84). This upper limit on  R is however compatible with what is expected from the 
observed value of the iron line equivalent width, since EW/R $\sim$100--130\,eV \citep{Perola:2002}.

The cross-calibration constant between \emph{XMM} and \emph{INTEGRAL/IBIS} is found to be 
1.48$^{+0.45}_{-0.34}$, suggesting again that the source is not strongly variable over 
long periods of time despite the previous \emph{INTEGRAL} indication; in fact, 
the 2--10\,keV flux measured during the 2013 pointing is 2$\times$10$^{-12}$erg cm$^{-2}$s$^{-1}$,
again consistent with the \emph{XMM} Slew and \emph{XRT} data. 

The total (2--100\,keV) X-ray luminosity is 2.7$\times$10$^{45}$erg\,s$^{-1}$, confirming that the 
source is extremely powerful at these high energies. Taking into account the information gathered 
for this source in the available bands, we can estimate that the bolometric luminosity exceeds at 
least 10$^{46}$erg\,s$^{-1}$. Indeed, if we evaluate the bolometric luminosity, assuming the 
relation (L$_{\rm Bol}$=25$\times$L$_{\rm 20-100keV}$) found by \citet{Mushotzky:2008}, we obtain
a value of 7$\times$10$^{46}$erg/s\footnote{The \emph{Swift/BAT} hard X-ray luminosity 
(14--195\,keV) has been rescaled to the \emph{INTEGRAL/IBIS} 20--100\,keV one.}
The Eddington ratio is therefore around 0.3, 
implying a highly efficient AGN at the centre of this galaxy.

As a final remark, we note that IGR J17488--2338 is somehow peculiar in the X-rays. 
At odds with the expectations of the unified theory, it is absorbed 
in the X-rays, but also displays broad emission lines in the optical;
furthermore, contrary to what is generally observed in radio galaxies where reprocessing 
features are often weak \citep{Grandi:2006}, it shows a prominent iron line and could have a 
significant reflection component. It is clearly a source that deserves more in-depth studies to 
understand its peculiarities and properties at these high energies.
  
\section{Radio Characteristics }
From the NVSS (Northern VLA Sky Survey) 1.4 GHz contours displaied in Figure~\ref{atel} 
and Figure~\ref{wise}, two compact regions of radio emission can be seen within
the lobes. These compact regions are typical of the hotspots seen in Fanaroff-Riley Type II 
(FR II; \citealt{fanaroff:1974})  radio galaxies. From the NVSS
image we measured the flux density of the various components by means of
the AIPS tasks IMFIT (core) and TVSTAT (lobes), and obtained a flux
density of 60.5$\pm$2.1\,mJy for the core, and F$_{\rm 1.4GHz}$=179.8$\pm$6.3\,mJy
and 157.3$\pm$5.5\,mJy for the northern and southern lobe respectively
(including the hot spots).
The total flux density we measured from the image is
F$_{\rm 1.4GHz}$=398.5$\pm$13.9\,mJy, consistent with the sum of the individual components.
Unfortunately, no pointed radio observations are available in the radio
archives for this radio galaxy, which prevents us even from general
considerations on its properties.
At 4.85\,GHz, the PMN survey flux limit of 
42\,mJy is used in the calculation. We should point
out that the angular resolution of the PMN survey is much larger than
that of NVSS (almost a factor of 7), hence any spectral estimate is
impossible. However, from the basis of the flux density values reported
here, we can safely conclude that the spectrum of the radio core is
flat, pointing at ongoing nuclear activity, and that the lobes are steep,
most likely as consequence of radiative aging, which causes the frequency
break in the spectrum to shift to lower frequencies
(see \citealt{blundell:2000}.)

The 1.4\,GHz total flux density of IGR J17488--2338 (398.5\,mJy) 
corresponds to a radio power of 6.8$\times$10$^{25}$ W\,Hz$^{-1}$, a value well above the
FRI/FRII break luminosity, and consistent with the lobe and hot spot
morphology, typical of FRII radio galaxies. The core fraction derived at
1.4\,GHz from NVSS is about 0.15. This low value is consistent with the
overall low core dominance for FRII radio galaxies (e.g. \citealt{lara:2001}).
All the above characteristics fully qualify IGR J17488--2338 as an 
FR II radio galaxy of intermediate power.

At the frequency of 1.4\,GHz and using a line profile across the two
radio lobes, we measure the separation between the
two most distant points in the lobes 
as 480\,arcsec. At the source redshift of z = 0.24 and assuming current 
cosmological parameters (H$_0$ = 71, $\Omega_{\rm M}$ = 0.270, $\Omega_{\rm vac}$ = 0.730) 
the observed separation corresponds to a scale of 3.760\,kpc/arcsec.
Therefore, IGR J17488--2338 has a projected linear extent of 1.4\,Mpc, 
sufficiently large for it to be classified as a GRG.
Despite a search for similar objects in the NVSS recently carried out by \citet{Solovyov:2011},
IGR J17488--2338 was not recognized as a giant radio galaxy, probably because of its low location 
on the galactic plane; therefore the \emph{INTEGRAL} detection
rediscovered this source and for the first time found that it has a giant structure.

\section{Conclusions}

We have uncovered an AGN with a large scale radio structure in the newly 
discovered \emph{INTEGRAL} source IGR J17488--2338. The source, which is clearly a FR II of 
intermediate power, has a linear size of 1.4\,Mpc, which fully qualifies it as a giant radio galaxy. 
The source is remarkable also for other properties, among which 
its X-ray absorption characteristics at odds with its type 1 classification and the presence of an 
iron line in the X-rays. 

It is still unclear what are the reasons or conditions that lead to the formation of giant radio 
galaxies. It could be special external conditions (such as the low density of the intergalactic 
medium) or uncommon internal properties of the source central engine
(like a high jet power or a long activity time). It is likely that none of the mentioned 
reasons is sufficient in itself and several conditions must be actually satisfied to provide the 
large scale structures seen in some radio galaxies.
In the particular case of IGR J17488--2338, the properties of the central AGN are quite exceptional, 
suggesting that it may be capable of producing a highly powerful jet or of maintaining the activity 
over a long period of time; either possibilities provide the conditions to form a large scale radio 
structure. The source is extremely bright in the X/gamma-rays: among a set of 25 radio galaxies 
detected so far by \emph{INTEGRAL}, this is the brightest object in the sample and also one of the 
most efficient accretors (Molina et al., in prep.).
Like Cygnus A and 4C 74.26, also included in the \emph{INTEGRAL} sample of radio galaxies,  
IGR J17488--2338 hosts a black hole with a mass greater than 10$^9$ solar masses; coupled with the 
source extension, this value perfectly fits with the linear relation found by \citet{Kuzmicz:2013} 
for GRG and based on the observed linear extensions and the black hole masses derived from the 
H$\alpha$ emission line. It is interesting to note that also 4C 74.26 is a
giant radio galaxy \citep{Ishwara-Chandra:1999} with an extension of 1.9\,Mpc, i.e. very similar to 
that of IGR J17488--2338. 
It is therefore possible that the hard X-ray selection made available by \emph{INTEGRAL} and /or 
\emph{Swift/BAT} allows the detection of the brightest AGN in the sky, and
consequently also of the most powerful radio galaxies, i.e. those that are able to produce large  
scale radio structures. Indeed, among the sample of \emph{INTEGRAL} detected radio galaxies, 
6 (or 24\%) qualify as giant radio galaxies; this fraction is higher than what is generally 
found using radio surveys, which report fractions in the range 6--11\%,
depending on the survey used (\citealt{laing:1983}, \citealt{saripalli:2012}). This suggests that hard 
X-ray observations can provide a much more efficient way to find giant radio galaxies than radio ones, 
at least in the local Universe. A complete analysis of \emph{INTEGRAL} radio galaxies 
is underway and the results regarding this issue will be presented in a forthcoming dedicated paper.
 
\begin{acknowledgements}

We acknowledge the Italian Space Agency (ASI) financial programmatic support via contract ASI/INAF 
n. 2013-025.R.O. 
This research has made use of the  HEASARC archive provided by NASA's Goddard Space Flight Center 
Italian Space. We thank Dr. R Landi for providing figure 1. 

\end{acknowledgements}

\bibliography{/Users/manu/Desktop/science/mol_biblio}

\end{document}